\documentclass[12pt]{article}
\usepackage[utf8]{inputenc}
\usepackage{geometry}
\usepackage{fancyhdr}
\usepackage{sectsty} 
\usepackage{graphicx} 

\usepackage{amsmath}
\usepackage{mathtools}
\usepackage{multicol}
\usepackage{amssymb}
\usepackage{multirow}
\usepackage{float}
\usepackage{afterpage}
\usepackage{cleveref}
\usepackage{xcolor}
\usepackage[normalem]{ulem}

\usepackage[labelfont=bf]{caption}

\usepackage[
maxcitenames=2,
backend=biber,
style=authoryear,
]{biblatex}
\DeclareNameAlias{sortname}{last-first}

\usepackage{caption} 

\usepackage{times}

\title{\textbf{\MakeUppercase {\Large{Towards Real-World Validation of a Physics-Based Ship Motion Prediction Model}}}}
\author{
    \normalsize{\textbf{Michail Mathioudakis}, DeepSea Technologies, m.mathoudakis@deepsea.ai}\\
    \normalsize{\textbf{Christos Papandreou}, DeepSea Technologies, c.papandreou@deepsea.ai}\\
    \normalsize{\textbf{Theodoros Stouraitis}, DeepSea Technologies, t.stouraitis@deepsea.ai}\\
    \normalsize{\textbf{Vicky Margari}, School of Naval Architecture and Marine Engineering,}\\ \normalsize{National Technical University of Athens, vickymargari@mail.ntua.gr}\\
    \normalsize{\textbf{Antonios Nikitakis}, DeepSea Technologies, a.nikitakis@deepsea.ai}\\
    \normalsize{\textbf{Stavros Paschalakis}, DeepSea Technologies, s.paschalakis@deepsea.ai}\\
    \normalsize{\textbf{Konstantinos Kyriakopoulos}, DeepSea Technologies, k.kyriakop@deepsea.ai}\\
    \normalsize{\textbf{Kostas J. Spyrou}, School of Naval Architecture and Marine Engineering,}\\ \normalsize{National Technical University of Athens, k.spyrou@central.ntua.gr}
}

\makeatletter
\renewcommand{\section}{\@startsection{section}{1}{\z@}%
                                   {-3.5ex \@plus -1ex \@minus -.2ex}%
                                   {2.3ex \@plus.2ex}%
                                   {\normalfont\bfseries\raggedright}}
\makeatother

\crefname{section}{Section}{Sections}
\newcommand{\dottedsection}[1]{%
  \refstepcounter{section}
  \section*{\arabic{section}. #1}
  \label{sec:#1}
}

\usepackage{titlesec}

\titleformat{\subsection}
  {\normalfont\fontfamily{phv}\fontsize{14}{17}}{\thesubsection}{1em}{\MakeUppercase}

\date{}
\usepackage{times}
\begin{document}

\renewcommand{\refname}{\normalsize REFERENCES}

\maketitle
\begin{center}
    \bfseries{ABSTRACT}
\end{center}
\noindent
The maritime industry aims towards a sustainable future, which requires significant improvements in operational efficiency. Current approaches focus on minimising fuel consumption and emissions through greater autonomy. Efficient and safe autonomous navigation requires high-fidelity ship motion models applicable to real-world conditions. Although physics-based ship motion models can predict ships' motion with sub-second resolution, their validation in real-world conditions is rarely found in the literature.

This study presents a physics-based 3D dynamics motion model that is tailored to a container-ship, and compares its predictions against real-world voyages. The model integrates vessel motion over time  and accounts for its hydrodynamic behavior under different environmental conditions. 
The model's predictions are evaluated against real vessel data both visually and using multiple distance measures. Both methodologies demonstrate that the model's predictions align closely with the real-world trajectories of the container-ship.

\begin{center}
    \bfseries{KEYWORDS}
\end{center}
\begin{center}
    Autonomous Vessels,
    Ship Dynamics Modeling,
    Sustainability,
    Hydrodynamic Behavior\\
\end{center}

\dottedsection{INTRODUCTION}
\vspace{-2mm}
The development of autonomous vessels is driven by the need to enhance operational safety, efficiency, and sustainability in the maritime industry. Autonomous technology can augment human capabilities to reduce human errors, which are a leading cause of maritime accidents. At the same time, autonomous vessels with precise navigation and route planning capabilities aim to optimise fuel consumption and minimize emissions.

To achieve these high-level aims of autonomy, it is essential for systems to be able to understand and predict the behavior of vessels under different conditions using motion models. The research community has a long history on developing maneuvering models of vessels (\cite{nomoto1957steering, motora1959measurement, abkowitz1964ship,fossen2011handbook}). These models simulate the dynamic interactions between a vessel and its environment, taking into account the hydrodynamics of the hull, the propeller and the rudder, as well as the influence of environmental conditions such as wind, waves, and currents. 
Such models are foundational components of autonomous decision support systems, which utilise them to optimise routes and make real-time adjustments to short-range plans. Their accuracy is critical to the systems' safety and efficiency, so robust frameworks for validating them against real-world data is imperative if greater autonomy is to be achieved. However, validations of vessel models against real-world full-scale ship data in real environmental conditions was seen to be scarce in literature. This study thus aims to make an initial step towards enhancing understanding of the real-world predictive capabilities of HD models, by presenting a framework for comparison of model predictions with real world data. 
 

The investigation focuses on the development and validation of a physics-based motion model for an 83-meter containership, the SUZAKU. The developed model integrates 3-Degrees of Freedom (DoF) dynamic equations, utilising a Hydrodynamics Derivative (HD) model. The forces equations are based on works by~\cite{spyrou2006asymmetric} and~\cite{kijima2002practical} to simulate surge, sway, and yaw motions. 
The models introduced by ~\cite{fujiwara1998estimation} and ~\cite{holtrop1982approximate} are also used for calculations of wind forces and water-vessel interaction coefficients. By leveraging real-world data from sea trials and operational measurements, the study aims to evaluate the model's accuracy and reliability, using both visual inspection and seven distance measures, with a view towards creating a benchmark for autonomous vessel development. 

This work contributes to maritime engineering by presenting a validation process towards obtaining physics-based motion models for real-world use. The reader
will gain insights into: \vspace{-2mm}
\begin{itemize}
    \item vessel motion modeling techniques, \vspace{-2mm}
    \item the utility of different distance measures for evaluating the accuracy of predictive models, \vspace{-2mm}
    \item the accuracy of an HD model in predicting the motion of a real-world vessel.
\end{itemize} \vspace{-2mm}
 
The rest of the paper is organized as follows. ~\cref{sec:related_Work} reviews previous work in the field of maneuvering models and their validation. The specifics on the maneuvering model developed to predict the motion of the vessel are then provided in ~\cref{sec:maneuvering_model}. ~\cref{sec:Evaluations} presents the results of the evaluation performed along with a comparison of different distance measures, while~\cref{sec:conclusions} outlines conclusions and future work directions. 
\vspace{-2mm}

\dottedsection{RELATED WORK}
\label{sec:related_Work}
\vspace{-2mm}
\subsection{Maneuvering Models}
\vspace{-2mm}
 Research on modern maneuvering models dates back to the 1940s, with the seminal work by~\cite{davidson1946turning}. Over the decades, significant contributions have been made;  ~\cite{motora1959measurement} and \cite{inoue1981hydrodynamic} explored the hydrodynamic aspects, while~\cite{motora1955course} and~\cite{nomoto1957steering} focused more on coarse stability principles. Studies on individual components, such as the hull, the propeller, and the rudder, have been instrumental in further refining these models.  
 
\cite{abkowitz1964ship} and~\cite{aastrom1976identification} utilized Taylor expansion with respect to state variables to derive a maneuvering model. This approach has proven effective in simulating ship dynamics and has been applied in various studies for control and prediction purposes. For example, ~\cite{norrbin1971theory} used the Abkowitz model to develop a framework for ship maneuvering analysis, and applied it to develop control algorithms for autonomous vessel navigation. 

The Maneuvering Modeling Group (MMG) model that provided a comprehensive framework representing the interaction of various ship components was developed by~\cite{ogawa1977mmg} and later enhanced by~\cite{yasukawa2015introduction}. In robotics, simplified vessel models have been used by~\cite{wang2018design} for demonstrating autonomous control in unmanned surface vehicles. The MMG model has been particularly influential due to its high accuracy in simulating maneuvering motions, attributed to its detailed hydrodynamic formulations. This model has been widely used in the design and validation of autonomous navigation algorithms. For instance, \cite{li2013active} developed a path-following control law using the MMG model, and~\cite{zhang2017ship} applied it to demonstrate a course-keeping control law. The MMG model has also been used in reinforcement learning frameworks for path-following control of unmanned surface vessels~\cite{zheng2022soft} as well as addressing autonomous berthing and unberthing problems~\cite{maki2020application,miyauchi2022optimization,suyama2022ship}.

In addition to the MMG model, several other models have been proposed towards calculating hydrodynamic forces during low-speed motions, especially with large drift angles. These include the Hydrodynamic Derivatives (HD) model~\cite{yasukawa2015introduction}, the Cross-Flow Drag (CD) model~\cite{yoshimura2012hydrodynamic}, and the Table (TBL) model~\cite{sutulo2015development}.  

The Hydrodynamic Derivatives (HD) model, which adds higher-order terms to conventional MMG derivatives, attempts to account for the complex hydrodynamic behavior of a vessel at large drift angles~\cite{kose1984mathematical, takashina1986ship, spyrou2006asymmetric, kijima2002practical}. While this approach can provide more detailed force calculations, it also introduces significant computational complexity. This complexity can sometimes lead to failures or inaccuracies at very low speeds, where the higher-order terms may not accurately represent the physical forces at play. Consequently, the HD model's utility is often limited in scenarios involving near-zero velocities, highlighting the need for more robust modeling approaches in such conditions. 

The Cross-Flow Drag (CD) model is based on the hydrodynamic characteristics of a ship's cross-section and it is applicable at low speeds, yet it can be challenging to parameterize accurately~\cite{ogawa1977mmg, oltmann1984simulation}. To address this challenge, \cite{yoshimura1988mathematical} proposed a simplified version of the CD model (SCD) by introducing certain hydrodynamic constants. \cite{yoshimura2009unified,yoshimura2012hydrodynamic} investigated the practical applications of this model in different vessel types and maneuvers. The model has been described as useful although there is no rational physical background on the hydrodynamic coefficients~\cite{okuda2023maneuvering}.

The Table (TBL) model, which uses captive model tests to obtain hydrodynamic forces, omits mathematical expressions by treating hydrodynamic coefficients as tabular values~\cite{okuda2023maneuvering, sutulo2015development}. This model is particularly robust at zero speed, making it useful for simulations involving large drift angles~\cite{delefortrie2016captive}.  

In recent years, black box models based on machine learning have also been explored. These include artificial neural networks~\cite{ moreira2003dynamic,rajesh2008system, zhang2013black,oskin2013neural,luo2016modeling,woo2018dynamic}, 
and Gaussian processes~\cite{arizaramirez2018nonparametric, xue2020system}. These models follow a different approach by learning from data rather than relying on physical principles, providing flexibility and adaptability to various operational scenarios.

For this study, an HD physics-based model was selected, as they are standard practice in the research community and have high prediction performance, especially in small drift angles. \vspace{-2mm}

\subsection{Model Validation}
\vspace{-2mm}
Validation of maneuvering models is crucial for ensuring their accuracy and reliability in predicting vessel behavior. Traditionally, many models have been validated against simulation data or model tests ~\cite{okuda2023maneuvering, yasukawa2015introduction} which, although of some use, may not always capture the full complexity of real-world maritime conditions. The reliance on simulation data can sometimes lead to discrepancies when these models are applied in real-world cases. Consequently, there is a noticeable gap in the literature regarding the validation of maneuvering models with real-world data. Few studies have undertaken the rigorous task of validating maneuvering models against actual vessel performance measures obtained from sea trials and operational data~\cite{jing2021analysis, maki2020application}. 

This paper aims to address this gap by providing validation of an HD physics-based model using real-world data from the SUZAKU vessel. It aims to contribute to the body of knowledge with empirical evidence, enhancing the confidence in the model's accuracy and its applicability to realistic maritime operations. This validation process not only strengthens the reliability of the model but also sets a precedent for future research in integrating real-world data into model validation practices.
\vspace{-2mm}

\begin{figure}[h!]
    \centering
    \includegraphics[width=0.9\linewidth]{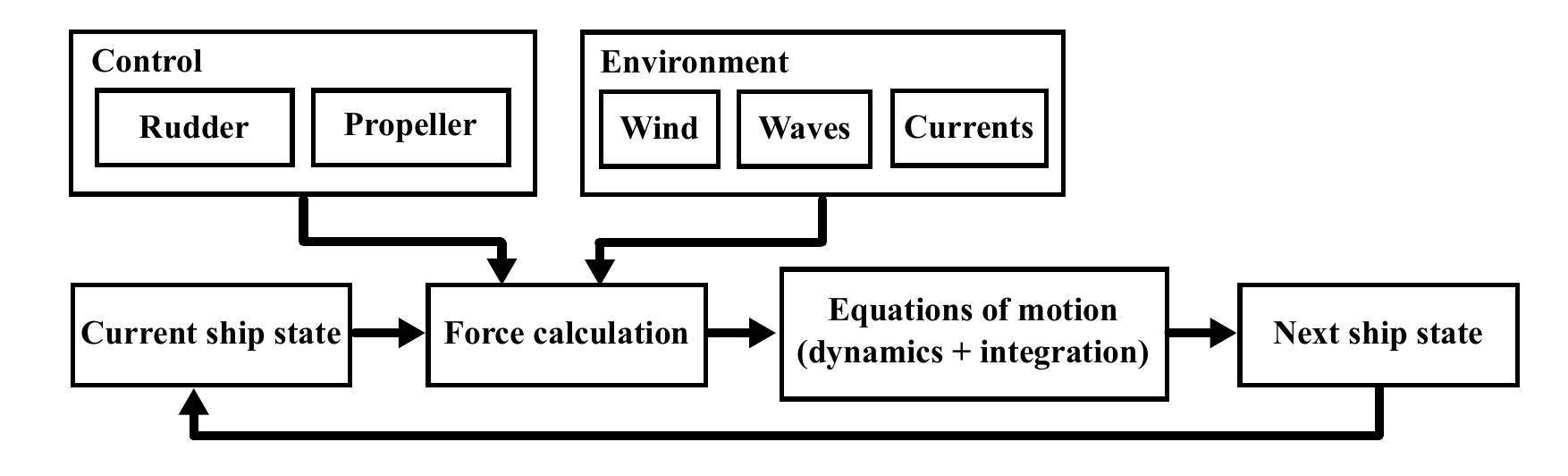} \vspace{-2mm}
    \caption{Computational flow of a physics-based vessel motion prediction model, with the following key blocks that are specific to ship maneuvering models: (a) ``Control": control commands of the rudder and the propeller, (b) ``Environment": environmental effects from the wind, waves, and sea currents, (c) ``Force calculation": computation of forced using the outputs of (a) and (b) along with hydrodynamic forces and the current ship state, and (d) ``Equations of motion": dynamics equations and integration over time to produce the next ship state.}
    \label{fig:simulation_system}
\end{figure}

\dottedsection{MANEUVERING MODEL}
\label{sec:maneuvering_model}
\vspace{-2mm}
The maneuvering model presented in this study, which hereafter is referred to as Full Analytical Model (FAM), is built on a physics-based representation and is designed to predict a vessel's movement over time. This model is utilised to forecast a vessel's trajectory with updates occurring at regular intervals (timesteps) as shown in~\Cref{fig:simulation_system}. Models of this nature enable the prediction of future positions and velocities based on current state and projected conditions. \vspace{-2mm}

\vspace{-2mm}

\noindent ~\Cref{fig:simulation_system} illustrates the key blocks of a maneuvering model, namely (i)
the \textit{force calculation}, which takes into account various forces acting on the vessel and (ii)  the \textit{dynamics equations}, along with the integration over time.
These components are utilized in order to predict the motion of the vessel following the step-by-step process described next. Given (i) the current state of the vessel, which includes its position, heading and velocities, (ii) 
the control commands considered, which include the rudder forces and the propeller thrust, and (iii) the environmental conditions such as sea currents, wind and waves, the \textit{force calculation} block, which also considers hydrodynamics, computes the 3D forces applied on the vessel at the current timestep. These forces along with the current state are fed into the \textit{dynamics equations} to compute the derivative of the state, which is integrated to compute the next state of the vessel.

In the rest of the subsections, we describe the specifics and details of the components of the maneuvering model, including the coordinate systems, the kinematics and dynamics, as well as the rudder, propeller and environmental force sub-models. \vspace{-2mm}
 


\subsection{Coordinate Systems}
\label{subsec:coordsys}
\begin{figure}[ht]
    \centering
    \begin{minipage}{0.4\textwidth}
        \includegraphics[width=0.95\linewidth]{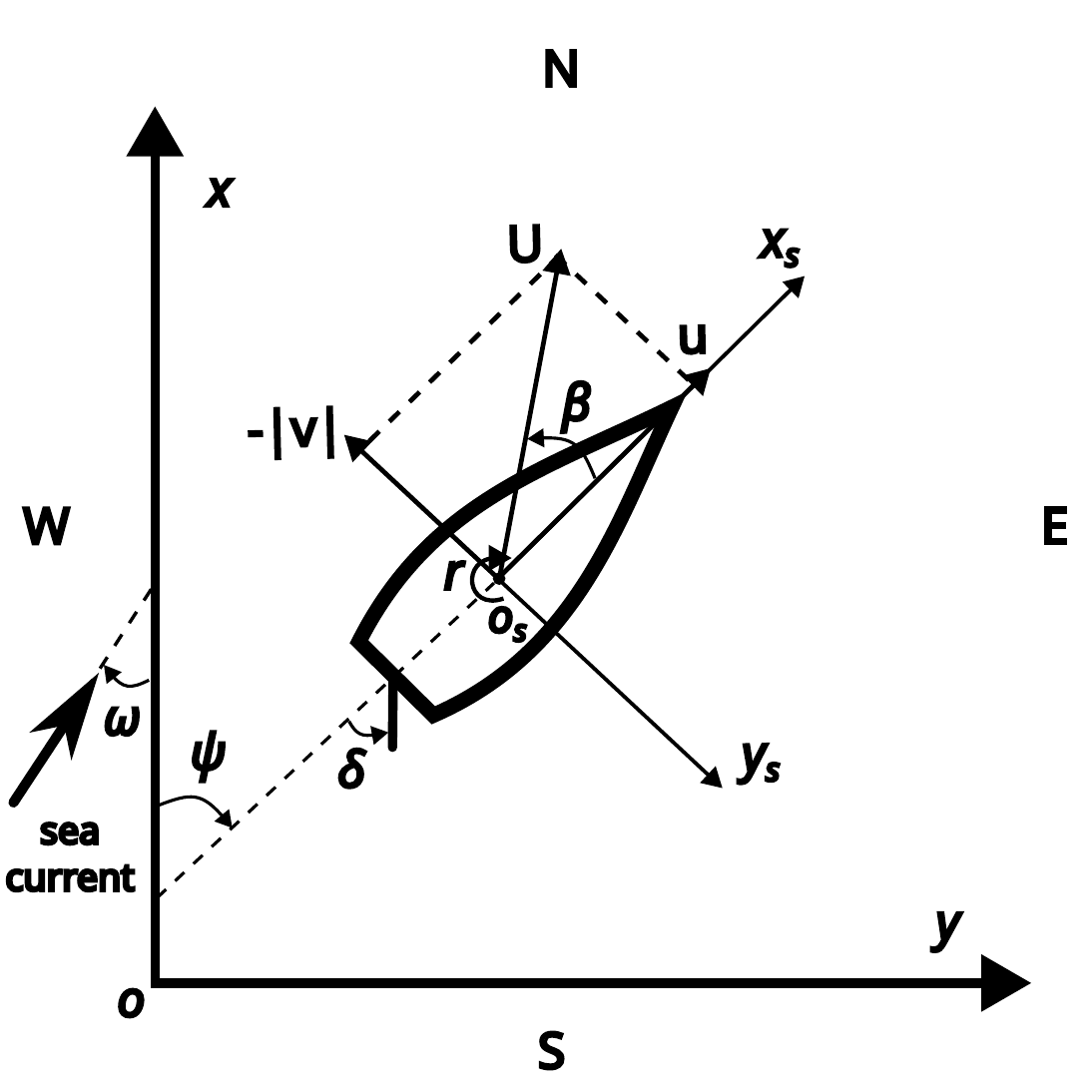}
    \end{minipage}\hspace{1cm} \vspace{1mm}
    \begin{minipage}{0.4\textwidth}
        \caption{The body-fixed coordinate system of the vessel is ${y_s-x_s}$, while the earth-fixed coordinate system is ${y-x}$. The position $O_S$ and heading $\psi$ of the vessel, and the direction of the sea current $\omega$ are depicted in terms of an earth-fixed coordinate system ${y-x}$. Surge speed $u$, sway speed $v$, total speed $U$ and yaw rate of turn $r$ are depicted in terms of a body-fixed coordinate system ${y_s-x_s}$. Also, the drift angle of the vessel is depicted as $\beta$ and the rudder angle as $\delta$.}
    \end{minipage}
    \label{fig:coord_system}
\end{figure}

\vspace{-2mm}
To describe the motion of the vessel, two horizontal systems of coordinates are used: one body-fixed (${x_s}$, ${y_s}$, ${z_s}$) and one earth-fixed (${x,y,z}$). Both are illustrated in~\Cref{fig:coord_system} and can be seen to be right-handed systems (${z}$ and ${z_s}$ are positive into the page).
\vspace{-2mm}


The origin, ${O_s}$ of the moving body-fixed system is located amidships, at the intersection of the longitudinal plane of symmetry with the undisturbed calm water
surface. The vessels’s trajectory is calculated for the earth-fixed system while the equations of motion (Dynamics, see~\cref{subsec:acc2vel}) are expressed with regards to the body-fixed system. The vessel's heading is denoted with $\psi$ with respect to the earth-fixed system. The surge, sway, and rate of turn are denoted with ${u}$, ${v}$ and ${r}$, while the ship's total speed is denoted with ${U} = \sqrt{u^2 + v^2}$. All the above speed terms refer to speed through water ($stw$).
The positive rudder angle $\delta$, translates to starboard rudder and $\beta$ is the drift angle, which are all defined with respect to the body-fixed system. \vspace{-2mm}

\subsection{Kinematics}
\label{subsec:vel2pos}
\vspace{-2mm}
The trajectory of the ship is obtained by integrating its velocity relative to the earth-fixed system $O$ over time. This velocity is given by its speeds over ground ($sog$) in the three earth-fixed dimensions. These are in turn derived by transforming the corresponding speeds $u'$, $v'$ and $r'$ from the body-fixed system $O_s$ by solving the following system: \vspace{-2mm}
\begin{equation}
    \begin{aligned}
        \label{eq:Trajectory equation 1}
        \dot{x} =  u' \cos{\psi} - v' \sin{\psi}\\
        \dot{y} =  u' \sin{\psi} + v' \cos{\psi}\\
        \dot{\psi} = r'
    \end{aligned}
\end{equation} 
Here, $u'$ and $v'$ represent surge and sway speeds over ground ($sog$). Yaw rate of turn over ground $r'$ is assumed equal to yaw rate of turn through water $r$. Hence, the transform above can be conveniently represented with a 2D rotation matrix that depends on $\psi$. \vspace{-2mm}


\subsection{Dynamics}
\label{subsec:acc2vel}
\vspace{-2mm}
To compute the speed through water ($stw$) versions of surge, sway and rate of turn ${u}$, ${v}$ and ${r}$, a 3-DoF nonlinear mathematical model from ~\cite{spyrou1996dynamic} is used.
This is an HD model (see~\cref{sec:related_Work}) described by the following differential equation: \vspace{-1mm}
\begin{equation}
    \label{eq:dynamics}
    \begin{split}
        &(m - X_{\dot{u}})  \dot{u} +  (Y_{\dot{v}} -X_{vr} -m) vr + (Y_{\dot{r}} - m x_G) r^2  = \Sigma X \\
        &(m - Y_{\dot{v}}) \dot{v} + (m x_G -  Y_{\dot{r}})\dot{r} + (-Y_{v}U v + (mu - Y_rU)r -Y_{vv} v |v| - Y_{vr} v |r| - Y_{rr} r |r|) = \Sigma Y  \\
        &(m x_G - N_{\dot{v}}) \dot{v} + (I_z - N_{\dot{r}})\dot{r} + (- N_v U v 
         + (m x_G u - N_rU) r  - N_{rr} r |r| - N_{rrv} \frac{rrv}{U} - N_{vvr} \frac{vvr}{U}) = \Sigma N 
    \end{split}
    \raisetag{3\normalbaselineskip}
\end{equation}   
where m is the mass of the ship, $I_{zz}$ is the yaw moment of inertia and $x_G$ is the longitudinal distance of ship’s centre of gravity from the moving axes’ origin, $O_s$. The $X$, $Y$ and $N$ terms with velocity ($u$, $v$, $r$) and acceleration ($\dot{u},\ \dot{v},\ \dot{r}$) subscripts are the maneuvering coefficients of the ship, and capture information regarding the hydrodynamic effects of its geometric characteristics. The hydrodynamic coefficients were calculated for the selected ship according to the methods presented in ~\cite{inoue1981hydrodynamic} and ~\cite{Clarke1982TheAO}. The left hand side of~\eqref{eq:dynamics} incorporates terms that express the added mass, damping and restoring forces phenomena. The external forces and moments acting on the ship are on the right side (RHS) and include those imposed from the rudder $[X_{rud}, Y_{rud}, N_{rud}]^T$, the propeller $X_{thr}$, the the wind $[X_{wind},  Y_{wind},  N_{wind}]^T$ and the waves $[X_{wave},  Y_{wave},  N_{wave}]^T$. In addition, the resistance of the vessel $R(u$) is also included in the RHS as part of $\Sigma X$. \vspace{-2mm}

\subsection{Rudder} 
\label{Rudder}
The rudder forces $[X_{rud}, Y_{rud}, N_{rud}]^T$ are calculated following~\cite{spyrou1996dynamic, kijima2002practical}. Given the rudder angle $\delta$, shown in~\Cref{fig:coord_system}, the rudder forces are given by: \vspace{-2mm}
\begin{equation}
    \begin{split}
        X_{rud} = -(1- t_R)F_N \sin{\delta}\\
        Y_{rud} = -(1+a_H) F_N \cos{\delta} \\
        N_{rud} = -(1+a_H) x_RF_N \cos{\delta}
    \end{split}
\end{equation} 
where $t_R$ is the thrust reduction coefficient calculated equal to $t_P$ as per ~\cite{holtrop1982approximate}, $a_H$ is the rudder-to-hull interaction coefficient and is set to 0.2, 
$x_R$ is the longitudinal
coordinate of the rudder’s center of lift ($x_R < 0$ for the given coordinates’ system, in SUZAKU is around -40 meters) and $F_N$ is the normal force on the rudder given by 
$ F_N = L\cos({a_R}) + D\sin({a_R})$,
where $a_R$ is the effective rudder inflow angle and $L$ and $D$, the rudder lift and drag forces, are given by $  L = 0.5 \rho C_L A_R U_R^2 \text{ and }
       D = 0.5 \rho C_D A_R U_R^2 $,
where $A_R$ is the area of the rudder, for SUZAKU is around 5.8 square meters, $\rho$ is the sea water density (1025 $kg/m^3$), $U_R$ is the effective inflow water speed at the rudder, $C_L$ and $C_D$ are the lift and drag coefficients for a generic rudder, which are expressed as a function of the effective rudder inflow angle $a_R$. \vspace{-2mm}




\subsection{Propeller}
\label{subsec:Propeller}
\vspace{-2mm}
The propeller is assumed to only produce force along the surge direction of the vessel, which is referred to as thrust $X_{thr}$. The respective thrust can be modelled per ~\cite{kijima2002practical,spyrou2006asymmetric} as:  \vspace{-1mm}
\begin{align}
    X_{thr} = (1-t_p)\frac{\rho n^2 D^4 k_T(J)}{1000}
\end{align}
while the torque $Q_{prop}$ demanded by the propeller can be calculated as: \vspace{-1mm}
\begin{align}
    Q_{prop} = \frac{\rho n^2 D^5 k_{Q}(J)}{1000}
\end{align}
\noindent where $t_P$ is the thrust deduction coefficient per ~\cite{holtrop1982approximate}, $n$ are the propeller’s revolutions in rounds-per-second ($RPS$), $D$ is the
propeller’s diameter in meters ($\sim 3m$ for SUZAKU), and $k_T(J)$ and $k_Q(J)$ are the propeller thrust and torque coefficients, as functions of the advance constant $J$, calculated by: \vspace{-1mm}
\begin{align}
    J = \frac{U\cos{\beta}(1-w_p)}{nD}
\end{align}
where $w_P$ is the effective propeller wake fraction, calculated as per ~\cite{holtrop1982approximate} and $\beta$ is the drift angle which is calculated by $ \beta = \arctan(-v/U) $ where $v$ is the sway speed and $U$ the total speed of the vessel.
Once $J$ is calculated the thrust and torque coefficients can be represented with one third order polynomial each as $   k_{T}(J) = c_{1}+c_2J+c_3J^2~~\text{and}~~k_{Q}(J) = a_{1}+a_2J+a_3J^2$. 
The equations for the $k_T$ and $k_Q$ are derived from the propeller open water characteristics tests. \vspace{-2mm}

\subsection{Environment}
\vspace{-2mm}
Concerning the environmental phenomena that affect the motion of the vessel, in our maneuvering model we consider the influence of wind, waves and sea currents. For the estimation of the wind forces, we utilize the Fujiwara wind method according to~\cite{fujiwara1998estimation}. The forces resulting from the wind are 
\vspace{-1mm}
\begin{equation}
    \begin{split}
    X_{wind} =  \frac{1}{2} \cdot C_X(\psi_{wind}) \rho_A A_F U_{wind}^2 \\
    Y_{wind} = \frac{1}{2} \cdot C_Y(\psi_{wind}) \rho_A A_L U_{wind}^2 \\
    N_{wind} = \frac{1}{2} \cdot C_N(\psi_{wind}) \rho_A A_L L_{OA} U_{wind}^2
    \end{split}
\end{equation}where $L_{OA}$ length overall, $A_F$ is the longitudinal projected area, $A_L$ is the lateral projected area, $\rho_A$ is the density of the air. $C_X, C_Y, C_N$ are the wind coefficients for SUZAKU, which were calculated in wind tunnel tests. $U_{wind}$ and $\psi_{wind}$ are the apparent wind speed and angle, respectively.

For the wave forces, we implemented the wave correction as per \texttt{ITTC, STAWAVE1} following~\cite{ittc2014speed}. The force along the x axis due to the waves is 
\vspace{-2mm}
 \begin{equation}
    X_{wave} = \frac{1}{16} \rho g H_{W1/3}^2 B \sqrt{\frac{B}{L_{\text{BWL}}}}
\end{equation}
where \textit{B} is the breadth of the ship, $H_{W1/3}$ is significant wave height, and 
$L_{BWL}$ is the length of the bow on the water line to 95\% of maximum beam.  
This model is restricted to wave directions that are $\pm 45^\circ$ off bow, therefore we assume $Y_{wave} = 0$ and $N_{wave} = 0$.

Regarding the influence of sea currents, these change the water flow around the vessel. Hence, they are not expressed as forces on the vessel; instead, they alter its $stw$ velocities. To model this, we construct a vector $ s_{sog} = [u', v']^T $ with the $sog$ velocities of the ship (see~\eqref{eq:Trajectory equation 1}) and a vector $s_{stw} = [u, v]^T $ with the $stw$ velocities of the ship (see~\eqref{eq:dynamics}). We also define a vector $s_{sc} \in \mathrm{R}^2$ with the sea current velocities with respect to the earth fixed $O$ coordinate system, as shown in~\Cref{fig:coord_system}. These three vectors assemble the following relationship $ s_{sog} = s_{stw} + R(\psi)^{-1} s_{sc}~\refstepcounter{equation}(\theequation)\label{eq:tide}$, 
where $R(\psi)$ is the inverse of the rotation matrix of the vessel and transforms the sea currents from $O$ (the earth-fixed) to the $Os$ (the body-fixed) coordinate system. \eqref{eq:tide} is used to: (i)
model the influence of sea currents on the vessel's motion and (ii) relate $sog$ velocities (see~\eqref{eq:Trajectory equation 1}) with $stw$ velocities (see~\eqref{eq:dynamics}) of the vessel.  Hence, it binds together~\eqref{eq:Trajectory equation 1} with~\eqref{eq:dynamics}. Also, note that sea currents are typically described via magnitude and direction (angle $\omega$), as shown in~\Cref{fig:coord_system}. Instead, we express sea currents as via a vector $s_{sc}$, which is a suitable representation for~\eqref{eq:tide}. \vspace{-2mm}

\dottedsection{EVALUATIONS}
\label{sec:Evaluations}
\vspace{-2mm}
This section describes experiments that were conducted to evaluate the predictive capabilities of the maneuvering model FAM presented in \cref{sec:maneuvering_model}. First, in order to validate the accuracy of FAM, its trajectories are compared against those generated by an MMG model~\cite{okuda2023maneuvering, suyama2024parameter} for the same maneuvers. Next, an investigation is presented into different distance measures that aim to quantify the difference between two trajectories of a vessel. The aim of this investigation is to identify a measure able to quantify the similarity between trajectories, as commonly done in other disciplines~\cite{tao2021comparative}, and go beyond qualitative visual inspections that are typically used to date. Finally, FAM's predictions are compared against data from real voyages of, both visually as well as using the most prominent distance measure.  \vspace{-2mm}

\subsection{Visual Validation Against MMG}
\vspace{-2mm}
To initially validate the accuracy and reliability of the FAM vessel motion model, it is used to replicate the prediction experiments from ~\cite{okuda2023maneuvering} and ~\cite{suyama2024parameter}, with the trajectories it generates compared visually against those generated by the MMG in the original studies. The trajectories reported in those studies (see Figure 3 in~\cite{suyama2024parameter}) were validated against comprehensive experimental scaled model trials in ~\cite{okuda2023maneuvering}. Thus, comparing FAM's performance against those trajectories serves as a robust first benchmark for assessing our model under the most realistic conditions previously available.

The experiments with the scaled vessel conducted by~\cite{okuda2023maneuvering} involved various maneuvers, including starboard and port turns, with specific initial speeds and RPM. By replicating these conditions, the trajectories generated by FAM can be directly compared to the empirically validated trajectories reported in these studies.

The specific maneuvers considered include a starboard turn with $106~RPM$ and a $+35^{\circ}$ rudder angle and an initial speed of $6$ knots, as well as a port turn with $106~RPM$ and a $-35^{\circ}$ rudder angle and an initial speed of $6$ knots. Those conditions were chosen because they represent challenging scenarios that test the model's ability to handle sharp turns and maintain accurate trajectories under varying directional changes. 

\begin{figure}[h!]
    \centering
    \includegraphics[width=0.85\linewidth]{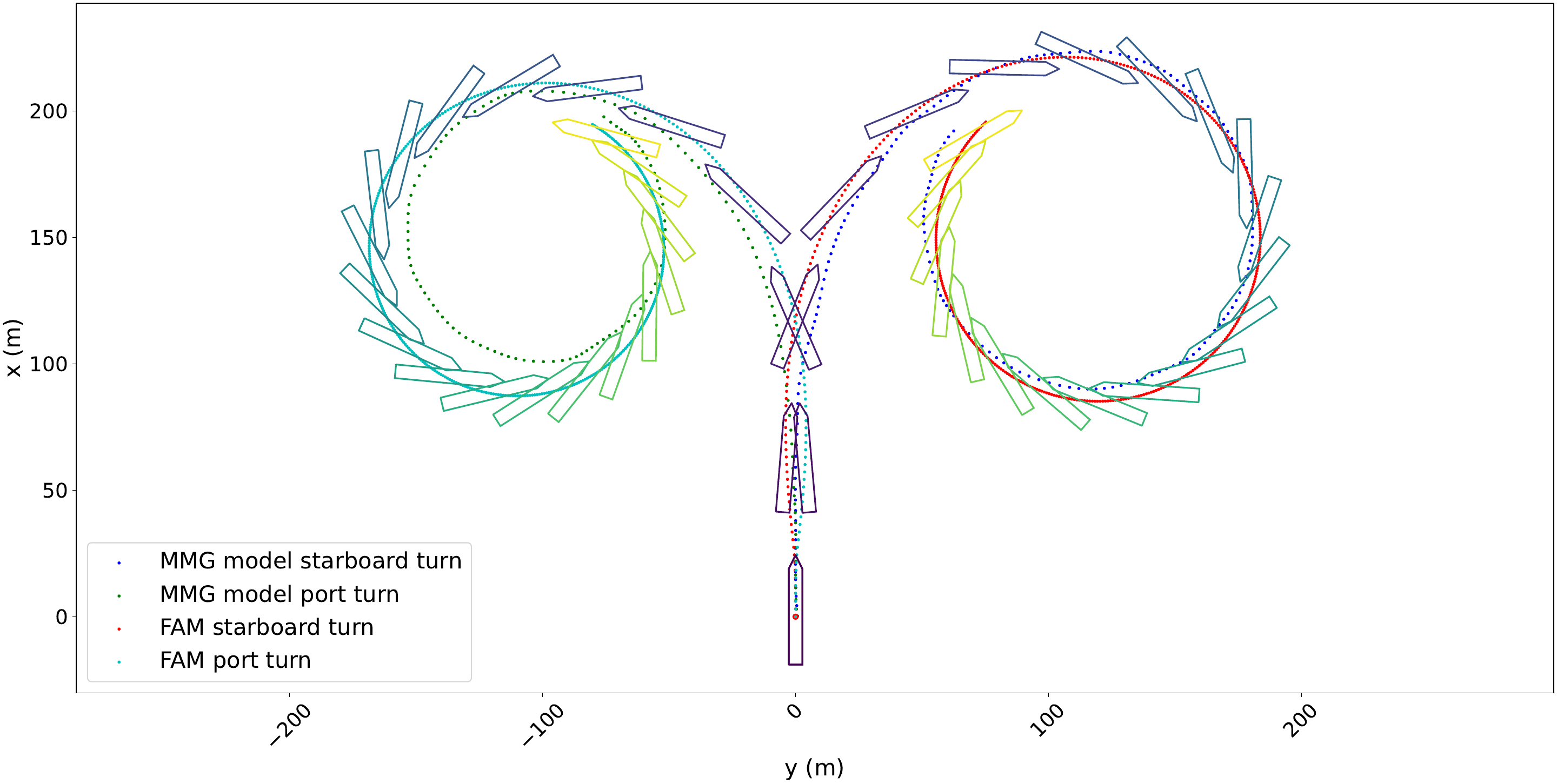} \vspace{-2mm}
    \caption{Visual comparison of trajectories from FAM (our model) and an MMG model as described and reported in (Figure 3 in~\cite{suyama2024parameter}).  \textit{N.B. The ship icons depict the heading of the FAM model. There is no actual trajectory of the real SUZAKU vessel in this plot, just a comparison between two models. The size of the vessel has been scaled for visual purposes and does not represent the actual size of the vessel.}} \vspace{-2mm}
    \label{fig:FAMvsMMG}
\end{figure}

\Cref{fig:FAMvsMMG} illustrates trajectories generated by FAM for the above mentioned maneuvers, showing a strong similarity with the results from~\cite{suyama2024parameter}. This indicates that our model effectively captures the essential dynamics of vessel movement. Further, in~\Cref{fig:FAMvsMMG} we can observe that the diameter of the starboard turning circle is larger than that of
the port turning circle, which matches the trajectories reported in~\cite{okuda2023maneuvering, suyama2024parameter}. This alignment further indicates the accuracy of the presented model, providing confidence in its use for predicting and optimizing vessel trajectories under various conditions, and  underscores the potential for practical applications of FAM in maritime navigation and control.

\vspace{-2mm}
\subsection{Distance measures for Quantitative Validation}
\vspace{-2mm}
To robustly validate motion models, it is necessary to move beyond visual inspection and develop a methodology of quantifying the similarity between trajectories. This subsection investigates number of distance measures that could serve this purpose. 

Seven distinct distance measures are tested across various different trajectories to identify the most representative. Measures are evaluated by their ability to encapsulate some or all of the following characteristics:  \vspace{-2mm} 
\begin{itemize}
    \item capturing distance deviation in $y-x$, as well as difference in heading, speeds and yaw rate of turn,  \vspace{-2mm}
    \item balancing the contribution of each dimension, to avoid bias in favour of one or more dimensions,  \vspace{-2mm}
    \item taking into consideration the specifics of each trajectory (average speed, total length of the trajectory etc.) e.g. a trajectory in maneuvering phase should not be penalised due to its low average speed.  \vspace{-2mm}
\end{itemize}
In essence, the primary goal is to ensure that the chosen distance measure accurately reflects the model's ability to generate trajectories that are as close as possible to the desired trajectories.

The distance measures investigated are listed below. For these distance measures, we define all dimensions with a bar to denote the true (ground truth) value of each dimension, \textit{e.g.} $\bar{x}$ denotes the value of $x$ value in the true trajectory. Further, we use the subscript $i$ to denote the knot points within the trajectory. This specifies the number of timesteps required to predict the state of the vessel, \textit{e.g.} $x_{10}$ indicates the value of $x$ for the vessel after $10$ timesteps. The total number of knot points is $n$. The distance measures are: \vspace{-2mm}

\begin{enumerate} 
    \item Mean Manhattan Distance (MMD): 
    $ MMD = \frac{1}{n} \sum_{i=1}^{n} \left| \bar{x_{i}} - x_{i} \right| + \left| \bar{y_{i}} - y_{i} \right|~\refstepcounter{equation}~~~~~~~~~~~~~~~~~~~~~~~~~~~~~~~~~~~~(\theequation)$
    
    \item Mean Euclidean Distance (MED):
    $ MED = \frac{1}{n} \sum_{i=1}^{n} \sqrt{(\bar{x_{i}} - x_{i})^2 + (\bar{y_{i}} - y_{i})^2}~\refstepcounter{equation}~~~~~~~~~~~~~~~~~~~~~~~~~~~~~~~~~~~~(\theequation)$
     
    \item Absolute State Distance (ASD):
    \begin{equation}
       ASD = \frac{1}{n} \sum_{i=1}^{n} \left( \left| x_i - \bar{x}_i \right| + \left| y_i - \bar{y}_i \right| + \left| \psi_i - \bar{\psi}_i \right| + \left| u_i - \bar{u}_i \right| + \left| v_i - \bar{v}_i \right| + \left| r_i - \bar{r}_i \right| \right)
    \end{equation}  
   
    \item Mean State Distance (MSD):
    \begin{equation}
       MSD =  \frac{1}{n} \sum_{i=1}^{n} \sqrt{(x_i - \bar{x}_i)^2 + (y_i - \bar{y}_i)^2 + (\psi_i - \bar{\psi}_i)^2 + (u_i - \bar{u}_i)^2 + (v_i - \bar{v}_i)^2 + (r_i - \bar{r}_i)^2}
    \end{equation}   
     
    \item Percentage of Manhattan Distance (PMD):
    \begin{equation}
       PMD = 100~\frac{1}{n} \sum_{i=1}^{n} \left( \frac{\left| \bar{x_{i}} - x_{i} \right| + \left| \bar{y_{i}} - y_{i} \right| + \left| \bar{\psi_{i}} - \psi_{i} \right| + \left| \bar{u_{i}} - u_{i} \right|}{ \bar{x_{i}}  +  \bar{y_{i}}  +  \bar{\psi_{i}}  +  \bar{u_{i}} } \right)
    \end{equation}  

    \item Percentage of Euclidean Distance (PED):
    \begin{equation}
        PED = 100~\frac{1}{n} \sum_{i=1}^{n} \left( \frac{\sqrt{(\bar{x_{i}} - x_{i})^2 + (\bar{y_{i}} - y_{i})^2 + (\bar{\psi_{i}} - \psi_{i})^2 + (\bar{u_{i}} - u_{i})^2}}{\sqrt{\bar{x_{i}}^2 + \bar{y_{i}}^2 + \bar{\psi_{i}}^2 + \bar{u_{i}}^2}} \right)
    \end{equation}  
    
    \item Custom Vessel Distance measure (cVDM):
    \begin{equation}
       \hspace{-1.5cm}
       cVDM = 100~\frac{1}{n} \sum_{i=1}^{n} \left( \frac{\left| \bar{x_{i}} - x_{i} \right|}{\bar{L}} + \frac{\left| \bar{y_{i}} - y_{i} \right|}{\bar{L}} + \frac{\left| \bar{\psi_{i}} - \psi_{i} \right|}{\pi} + \frac{\left| \bar{u_{i}} - u_{i} \right|}{\bar{U_{mean}}} + \frac{\left| \bar{v_{i}} - v_{i} \right|}{\bar{U_{mean}}} + \frac{\left| \bar{r_{i}} - r_{i} \right|}{r_{max}} \right)\hspace{-0.5cm}
    \end{equation} 

\end{enumerate} 
where $\bar{L}$ is the total length of the desired trajectory and $\bar{U_{mean}}$ is the average total speed of the desired trajectory. For SUZAKU, $r_{max}$ is set to \(0.0314\) radians per second (equivalent to 1.8 degrees per second), as it represents the maximum turning capability of the particular ship.

$MED$ and $MMD$ can be considered naive as they only take into account discrepancies along $x$ and $y$. They are considered the quantitative representation of visual inspection. They are not expressed as a percentage and their units is meters (m). Yet, we include them in this investigation as a useful baseline. The $ASD$ measure is an extension of the $MMD$, yet it takes into account all dimensions of the state ($x$, $y$, $\psi$, $u$, $v$, $r$). The same applies to the $MSD$ measure which is an extension of the $MED$, again considering all state dimensions. Both $MSD$ and $ASD$ are not expected to be consistent or understandable since they blend dimensions with different units, yet they are included for completeness reasons.

On the other hand, the rest of the measures studied, namely $PED$, $PMD$ and $cVDM$, have been calculated as percentages, hence the blending of different dimensions is expected to be better understood by humans. With regards to $PED$ and $PMD$ apart from $x$ and $y$ they also utilise $\psi$ and $u$. As a result we expect that they are more robust in  future deviations of the trajectories. Since they are percentages there is no issue adding together dimensions with originally different units. It should be noted, however, that $PED$, $PMD$ due to their design, they prone to balance issues. This means that if one of the ($x$, $y$, $\psi$, $u$,) is way greater than the rest, the total error will be unbalanced and lean towards the greater dimension. 

Last but not least, $cVDM$ is designed in a way, with all dimensions in place, which makes it relatively objective, \textit{e.g.} division of yaw rate of turn by the maximum turning capacity of the vessel. It is also adaptable so as to be able to take into account the \textit{e.g.} average total speed of the trajectory and not to penalise low speed maneuvers. It also considers the total length of the trajectory for $n = 120$ timesteps, in terms of meters, and as a result trajectories with short distance due to turning or maneuvers are not penalised. 

\begin{figure}[t]
    \centering
    \includegraphics[width=0.65\linewidth]{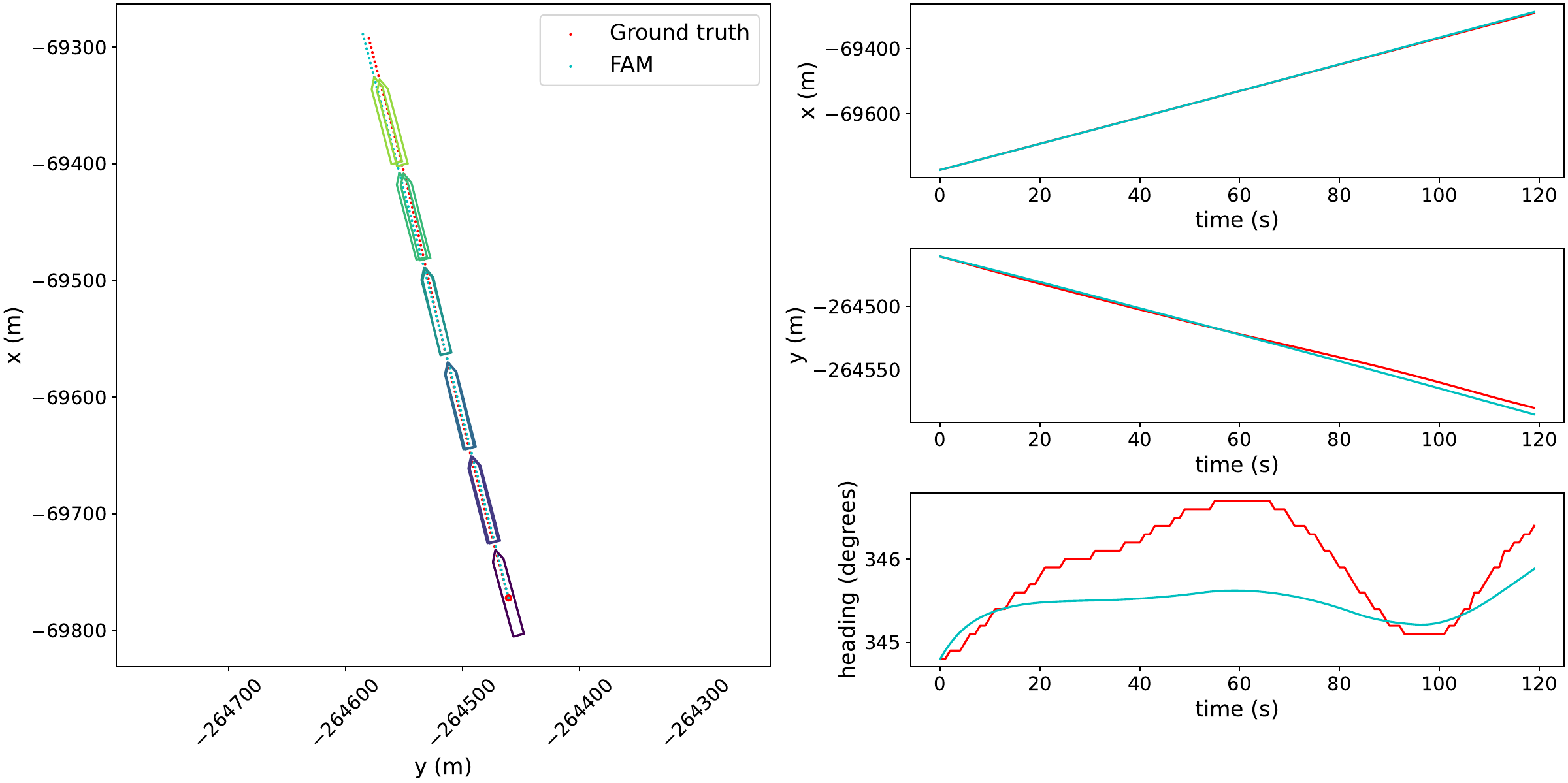}
    \caption{Error explanatory scenarios. Case Study 1. All dimensions are very close between actual trajectory and FAM.}  
    \label{fig:measures_example_1}
\end{figure}  

\begin{figure}[t]
    \centering
    \includegraphics[width=0.65\linewidth]{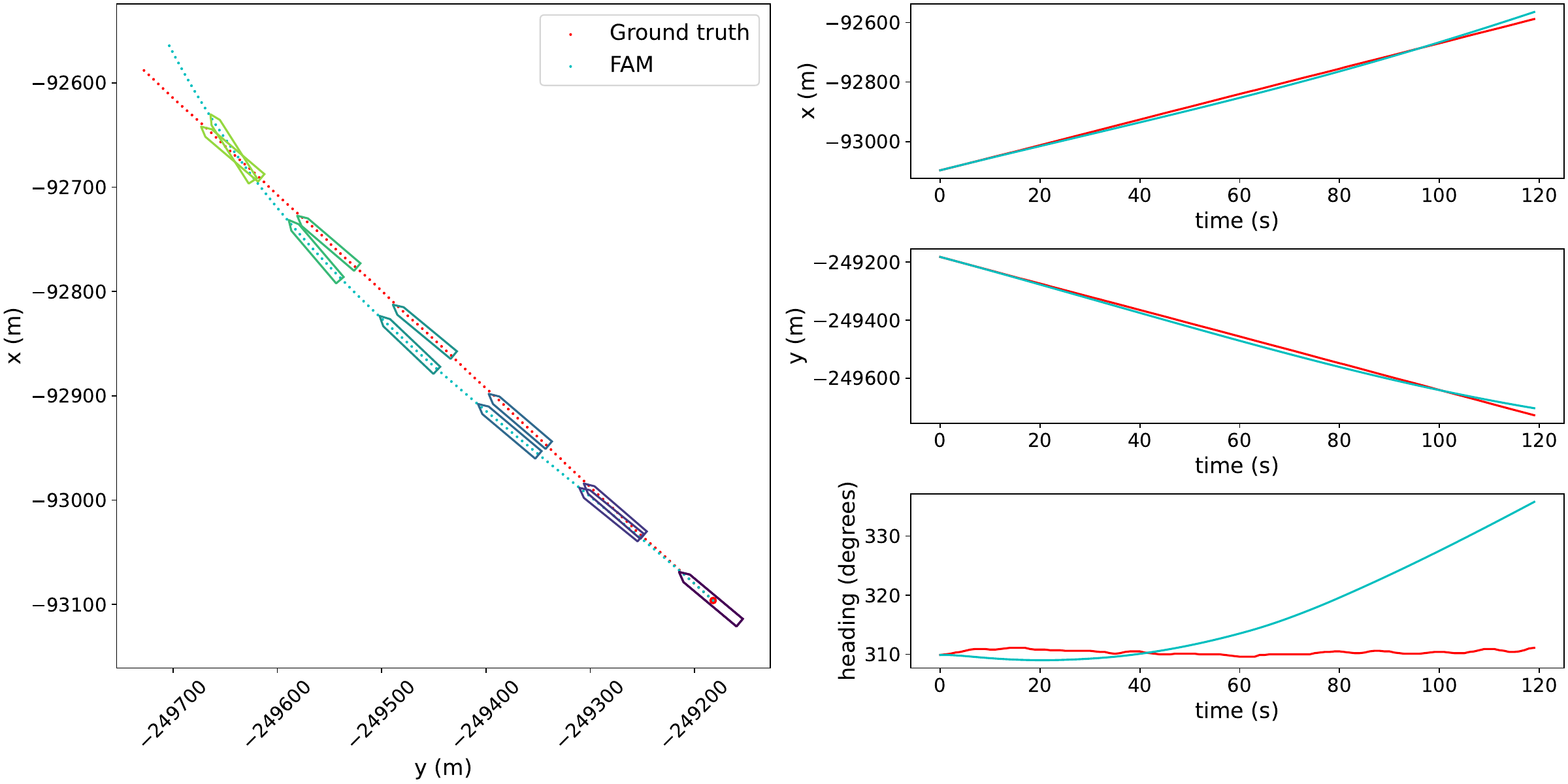}
    \caption{Error explanatory scenarios. Case Study 2. In this trajectory there is close distance in terms of $x, y$ but not as good in terms of heading $\psi$ and yaw rate of turn $r$.}  \vspace{-2mm}
    \label{fig:measures_example_2}
\end{figure}

\begin{figure}[t]
    \centering
    \includegraphics[width=0.65\linewidth]{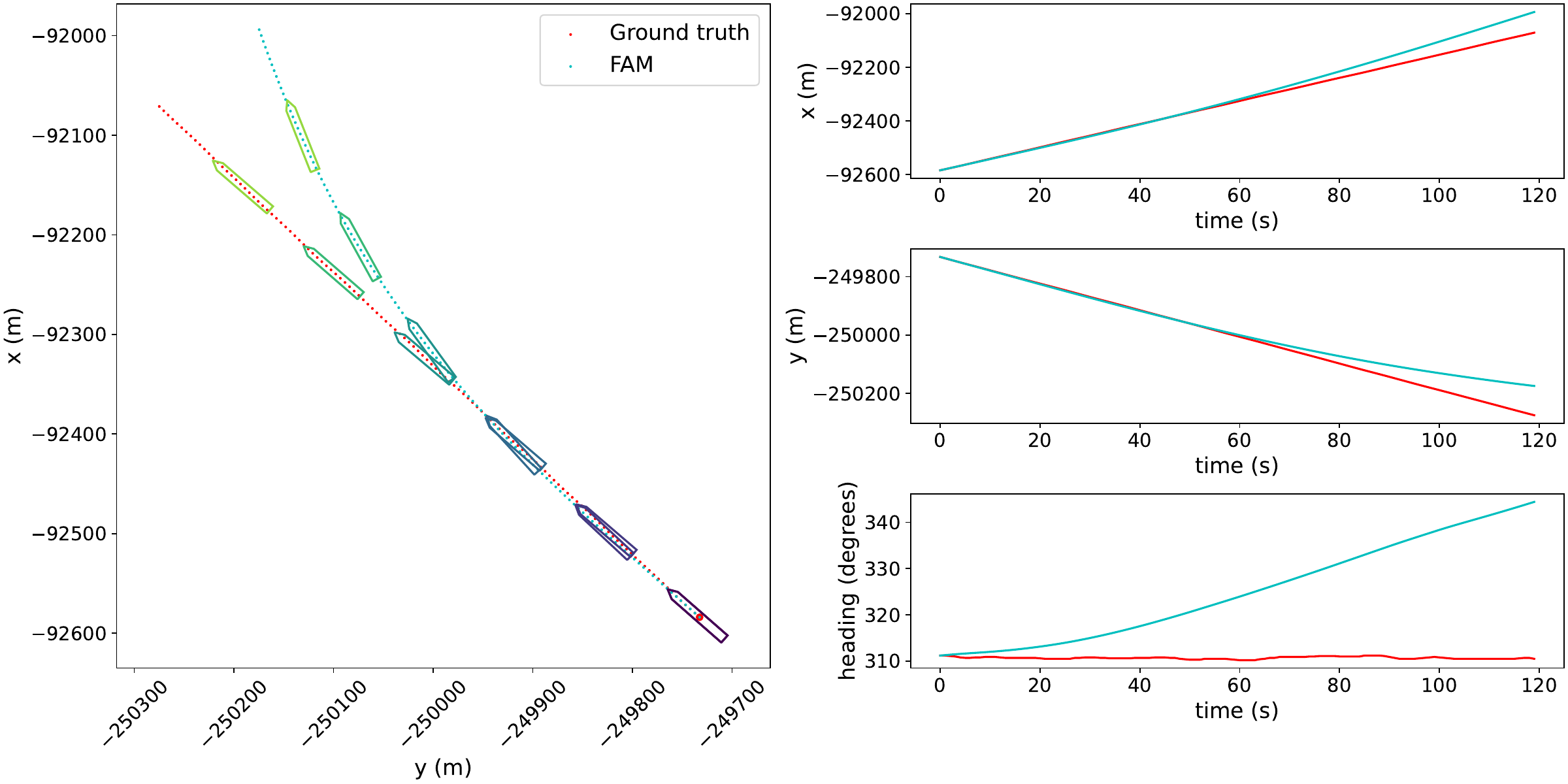}
    \caption{Error explanatory scenarios. Case Study 3. All dimensions are very close in the beginning of the trajectory which is not the case at the end.}  
    \label{fig:measures_example_3}
\end{figure}

\begin{figure}[!h]
    \centering
    \includegraphics[width=0.65\linewidth]{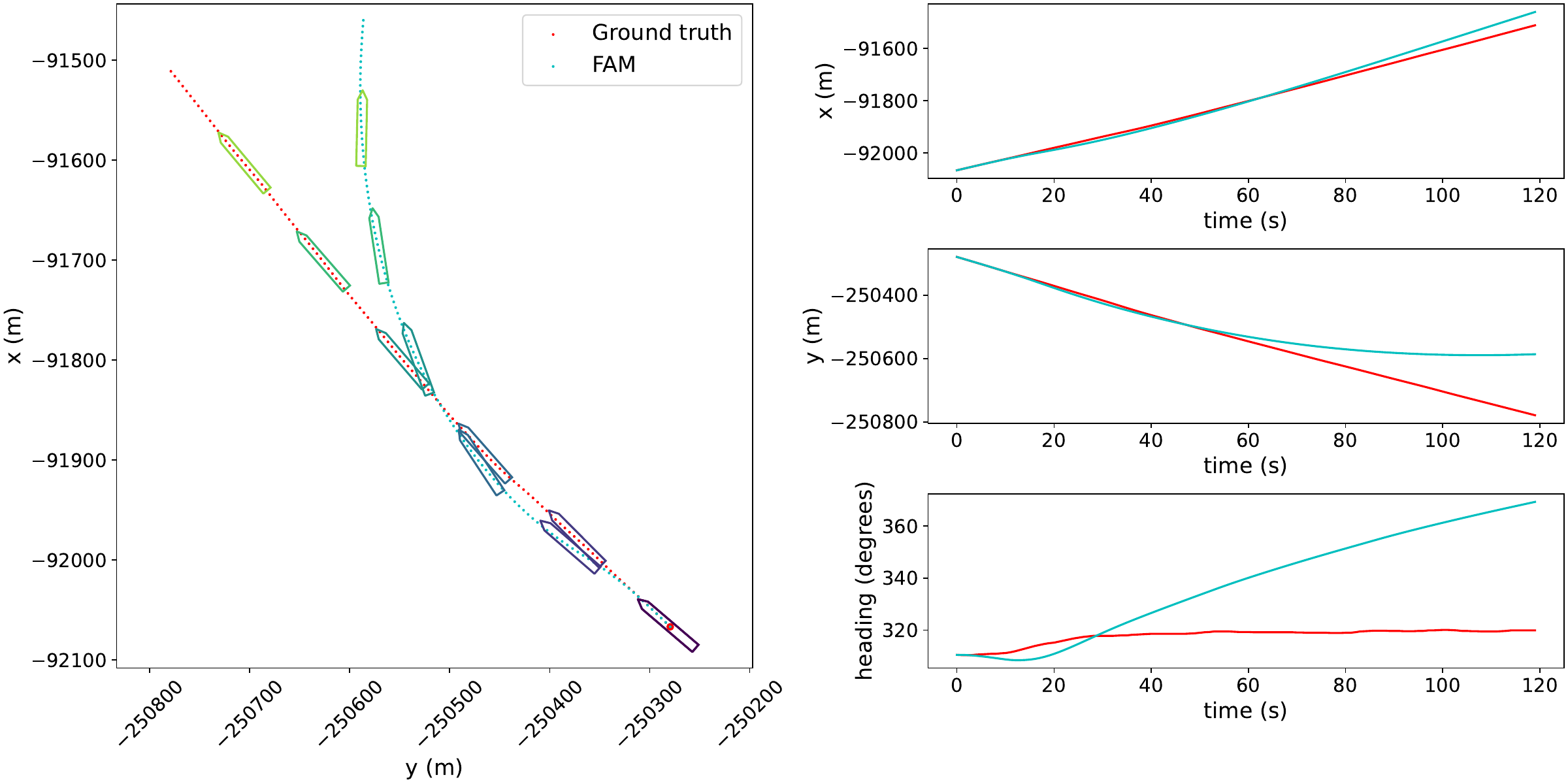}
    \caption{Error explanatory scenarios. Case Study 4. FAM diverged from the ground truth trajectory, as a result the differences in all dimensions are large.}  \vspace{-2mm}
    \label{fig:measures_example_4}
\end{figure}

In~\Cref{fig:measures_example_1,,fig:measures_example_2,,fig:measures_example_3,,fig:measures_example_4}, we present four arbitrary examples with $n = 120$ and $dt = 1 s$ that we utilize to compare ground truth trajectories with the trajectories predicted by FAM. Each figure includes a spatial illustration of the contrasted trajectories
along with tree one-dimensional plots, where the difference between the two trajectories can be inspected for each one of $x$, $y$ and $\psi$ independently. For each one of these examples, we applied the distance measures discussed above and we report with~\Cref{table:errors} the corresponding values. These  provide a clear and concise comparison from which we can observe the following.

Some of the common distance measures, \textit{e.g.} $MED$ and $MMD$ do not capture the discrepancy in the heading or the yaw rate of turn of the vessel. Yet, heading of the vessel is crucial for the following timesteps, as different heading will result in diverging trajectories. Prominent examples of this, are Case Study 2 (see~\Cref{fig:measures_example_2}) and Case Study 3 (see~\Cref{fig:measures_example_3}). If we use either the $MED$ or the $MMD$ measures it appears that the compared trajectories of Case Study 2 are a lot more similar than the compared trajectories of Case Study 3. However, we can observe that the heading in Case Study 2 is way off (together with the yaw rate of turn) and soon after it will result in divergent trajectories. This intricacy is effectively captured by the $PED$, $PMD$ and $cVDM$ distance measures which produce relatively close values in both case studies. The same applies for $ASD$ and $MSD$ measures as they cannot capture this intricacy as well, although they utilize both heading and yaw rate of turn.

\begin{table}[!h]
\caption{Summary of distance measures for the four different case studies shown in~\Cref{fig:measures_example_1,,fig:measures_example_2,,fig:measures_example_3,,fig:measures_example_4}. }
\centering
\begin{tabular}{|c|c|c|c|c|c|c|c|}
\hline
Case study &  MMD (m) & MED (m) & ASD & MSD  & PMD (\%) & PED (\%) & cVDM (\%) \\ \hline
1 & 3.0 & 2.4 & 3.1 & 2.5 & 1.6 & 1.5 & 1.0 \\ \hline
2 & 16.9 & 12.1 & 17.5 & 15.4 &  3.1 & 2.7 & 4.5 \\ \hline
3 & 42.6 & 30.2 & 43.3 & 34.6 &  5.4 & 4.8 & 6.3 \\ \hline
4 & 61.5 & 50.0 & 62.6 & 55.7 & 10.7 & 8.0 & 9.9 \\ \hline
\end{tabular}
\label{table:errors}
\end{table}


With regards to Case Study 1 (see~\Cref{fig:measures_example_1}) and Case Study 4 (see~\Cref{fig:measures_example_4}) we observe that all error measures can be used to quantify the distance between the trajectories. This is because in these scenarios the two contrasted trajectories are either very 'close' or very 'far'. Hence, the discrepancy is uniformly distributed across all dimensions. Further, by contrasting the values of $PED$, $PMD$ and $cVDM$ for Case Study 1 (see~\Cref{fig:measures_example_1}) and Case Study 2 (see~\Cref{fig:measures_example_2}), we can observe that $cVDM$ appears to be the most suitable for capturing the difference between identical trajectories (see~\Cref{fig:measures_example_1}) and trajectories that are similar but still different (see~\Cref{fig:measures_example_2}).


\begin{table}[!h]
\caption{Summary of custom Vessel Distance measure (cVDM) for the real-world scenarios under study.}
\centering
\begin{tabular}{|c|c|c|c|c|c|}
\hline
\multicolumn{6}{|c|}{\textbf{cVDM (custom Vessel Distance measure)}} \\ \hline
Optimal 1 & Optimal 2 & Satisfactory 1 & Satisfactory 2 & Sub-optimal 1 & Sub-optimal 2 \\ \hline
0.8 & 0.9 & 1.7 & 2.3 & 5.9 & 8.7 \\ \hline
\end{tabular}
\label{table:errors_real_world}
\end{table}

To conclude, the $cVDM$ measure, which incorporates not only positional coordinates but also the vessel's heading, surge speed, sway speed, and yaw rate of turn, offers a holistic and well balanced distance measure between two vessel trajectories. However, although $cVDM$ measure performs very well, this does not mean that will be the only measure that is going to be utilized. We strongly believe that using more than one measures provides diversification and can have a beneficial effect especially in niche scenarios. \vspace{-2mm}


\subsection{Validation against Real-World Voyages}
\vspace{-2mm}
Having studied the validity of different distance measures and demonstrated consistency with established literature, the next crucial step is to test the predictive capabilities of FAM on real-world data. This involves using actual vessel data and trajectories to evaluate the model's performance in practical scenarios. Real-world testing provides the most rigorous benchmark, incorporating a wide range of variables such as varying weather conditions (wind, waves, sea currents), different levels of hull fouling, and potential sensor inaccuracies. Successfully navigating these complexities will validate a model's robustness and its predictive capability to generalize across different situations.

\begin{figure} [t]
    \centering
    \includegraphics[width=0.65\linewidth]{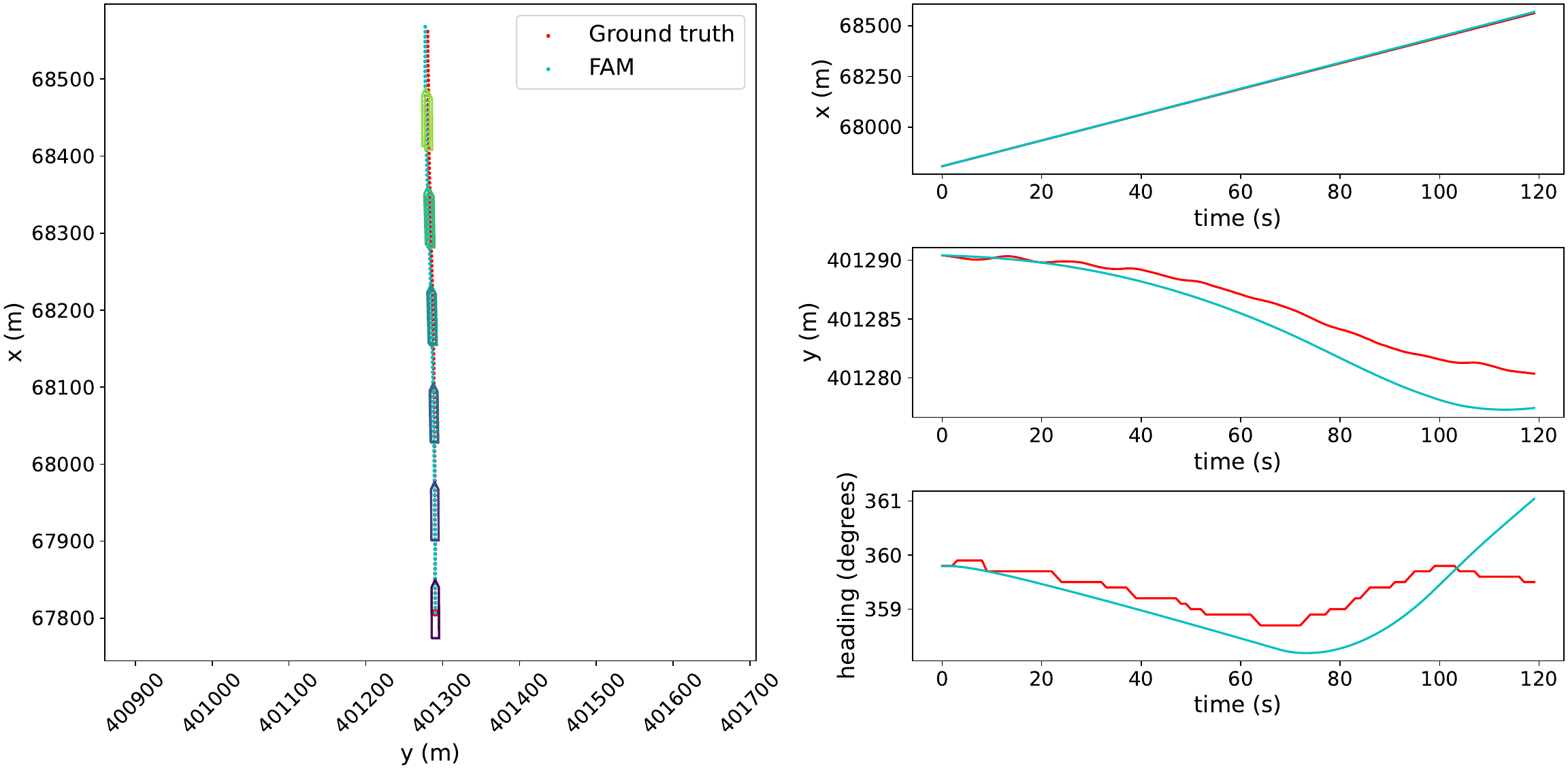}
    \caption{Comparison of FAM vs Real world data. Optimal scenario 1. All dimensions of FAM follow closely the actual trajectory.}  
    \label{fig:good1}
\end{figure}  

In terms of the weather conditions, such as wind, wave, and sea currents, the model considers the interaction of these factors with the ship’s geometry (e.g., beam and draft) to simulate the vessel's response in various sea states. To ensure that the environmental inputs reflect actual conditions experienced during the voyages, wind, wave and sea currents data were gathered for each voyage using hindcast weather data from the closest weather station. In addition, wind hindcast data samples were contrasted with on-board measurements of the anemometer to validate the conformity between the two. This ensures that FAM is tested in realistic operational environments, where the combined effects of weather can significantly impact vessel motion and trajectory.

\begin{figure}
    \centering
    \includegraphics[width=0.65\linewidth]{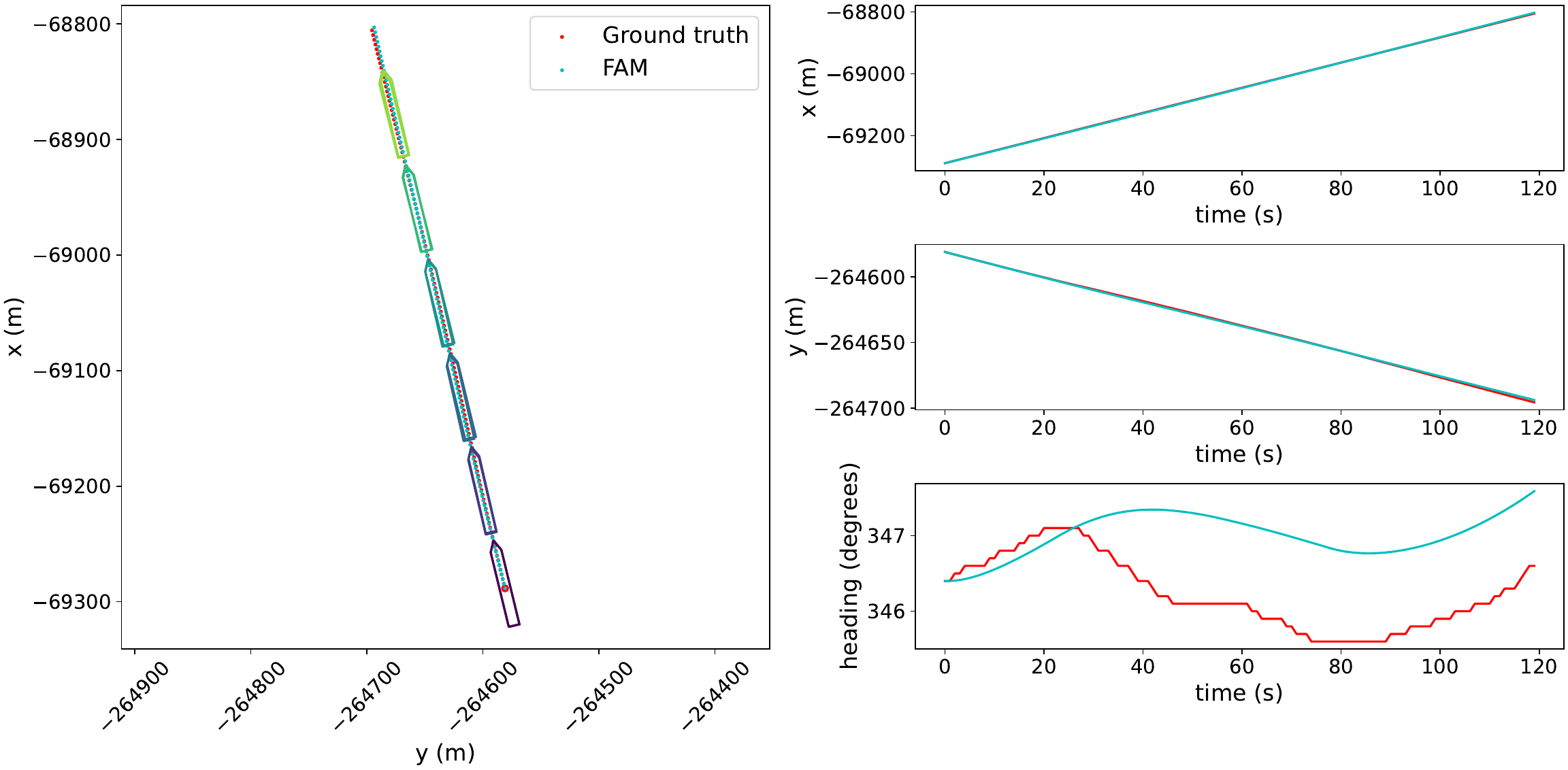} \vspace{-2mm}
    \caption{Comparison of FAM vs Real world data. Optimal scenario 2. FAM follows closely the actual trajectory in all measures.}
    \label{fig:good2}
\end{figure}

\begin{figure}
    \centering
    \includegraphics[width=0.65\linewidth]{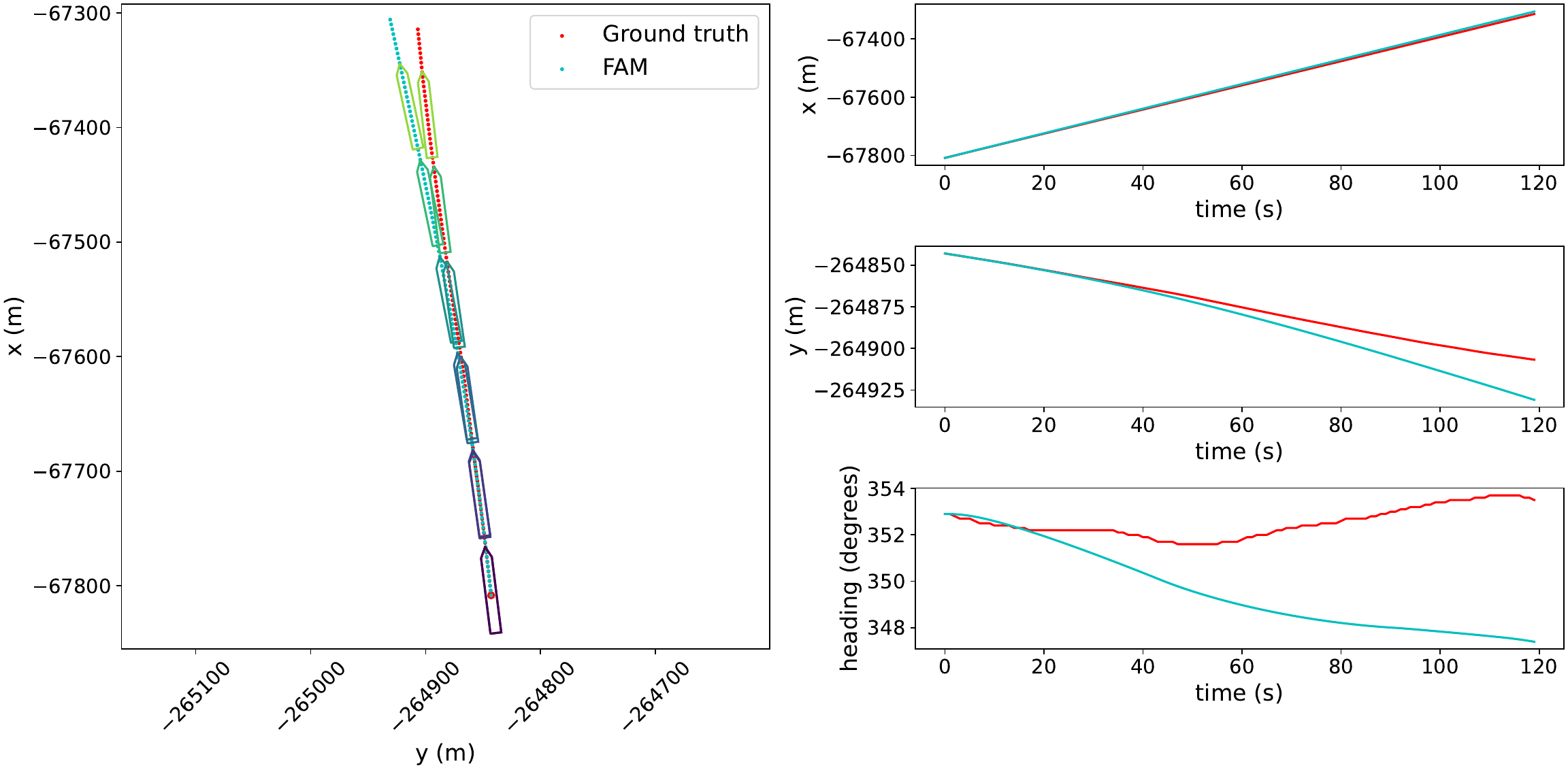}
    \caption{Comparison of FAM vs Real world data. Satisfactory scenario 1. FAM follows actual trajectory for half of the distance but then becomes sub-optimal.}
    \label{fig:good3}
\end{figure}

\begin{figure}
    \centering
    \includegraphics[width=0.65\linewidth]{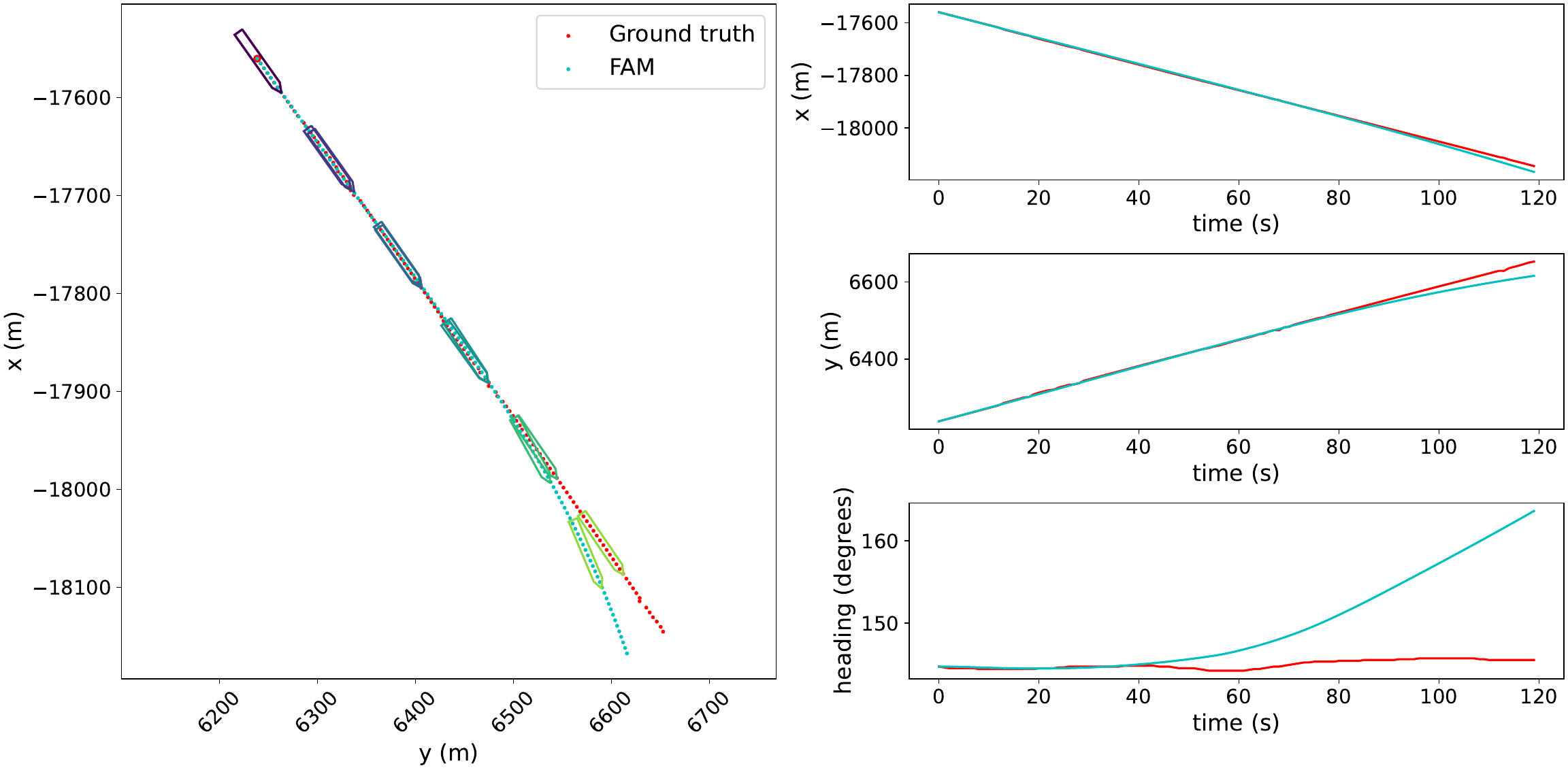}
    \caption{Comparison of FAM vs Real world data. Satisfactory scenario 2. FAM and actual data are close to each other for the bigger part of the trajectory but the deviation subsequently becomes large.}
    \label{fig:good4}
\end{figure}

\begin{figure}
    \centering
    \includegraphics[width=0.65\linewidth]{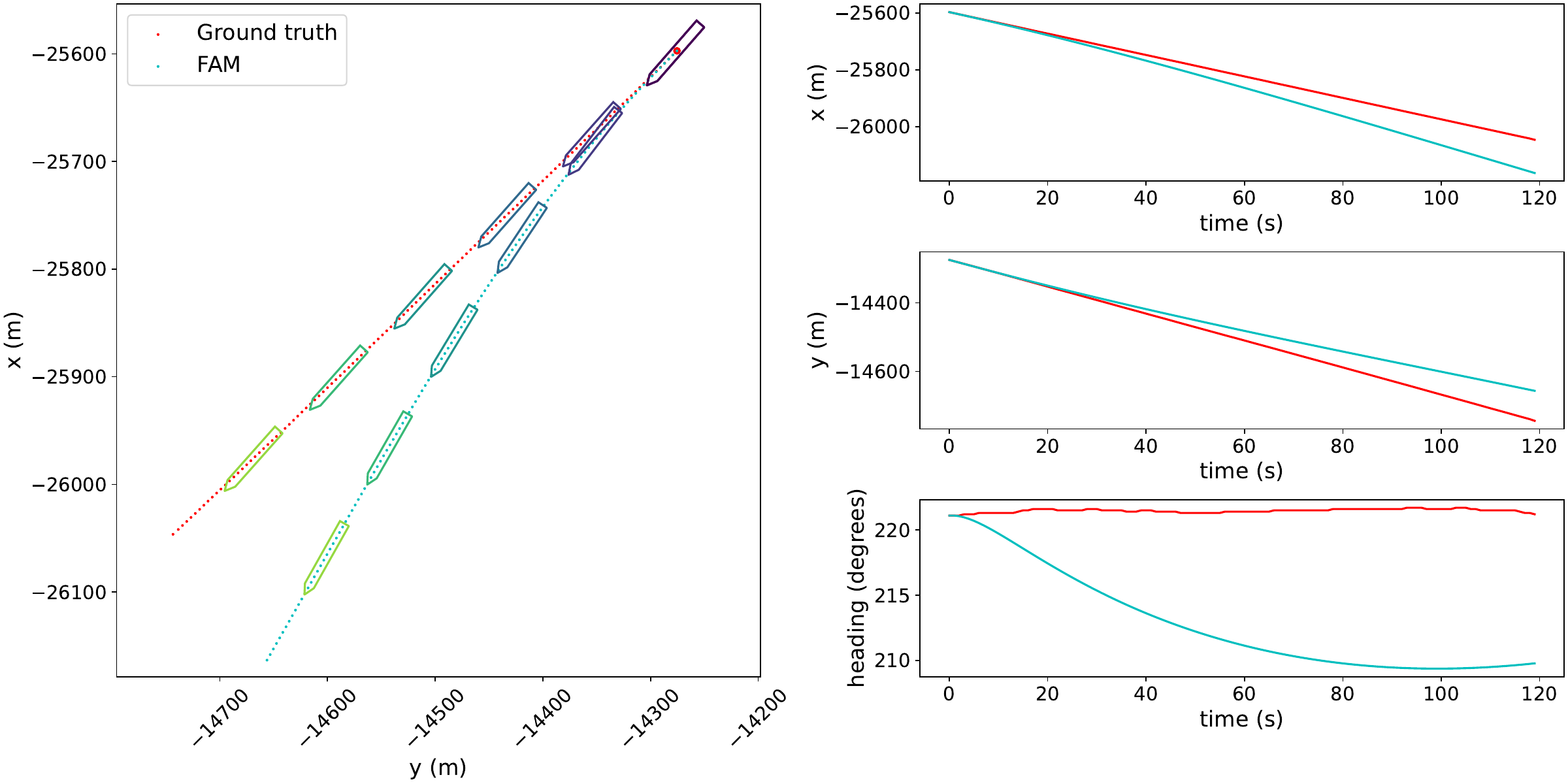}
    \caption{Comparison of FAM vs Real world data. Sub-optimal scenario 1. The difference between FAM and the actual trajectory grows large from the beginning of the trajectory.}
    \label{fig:good5} 
\end{figure}

\begin{figure}
    \centering
    \includegraphics[width=0.65\linewidth]{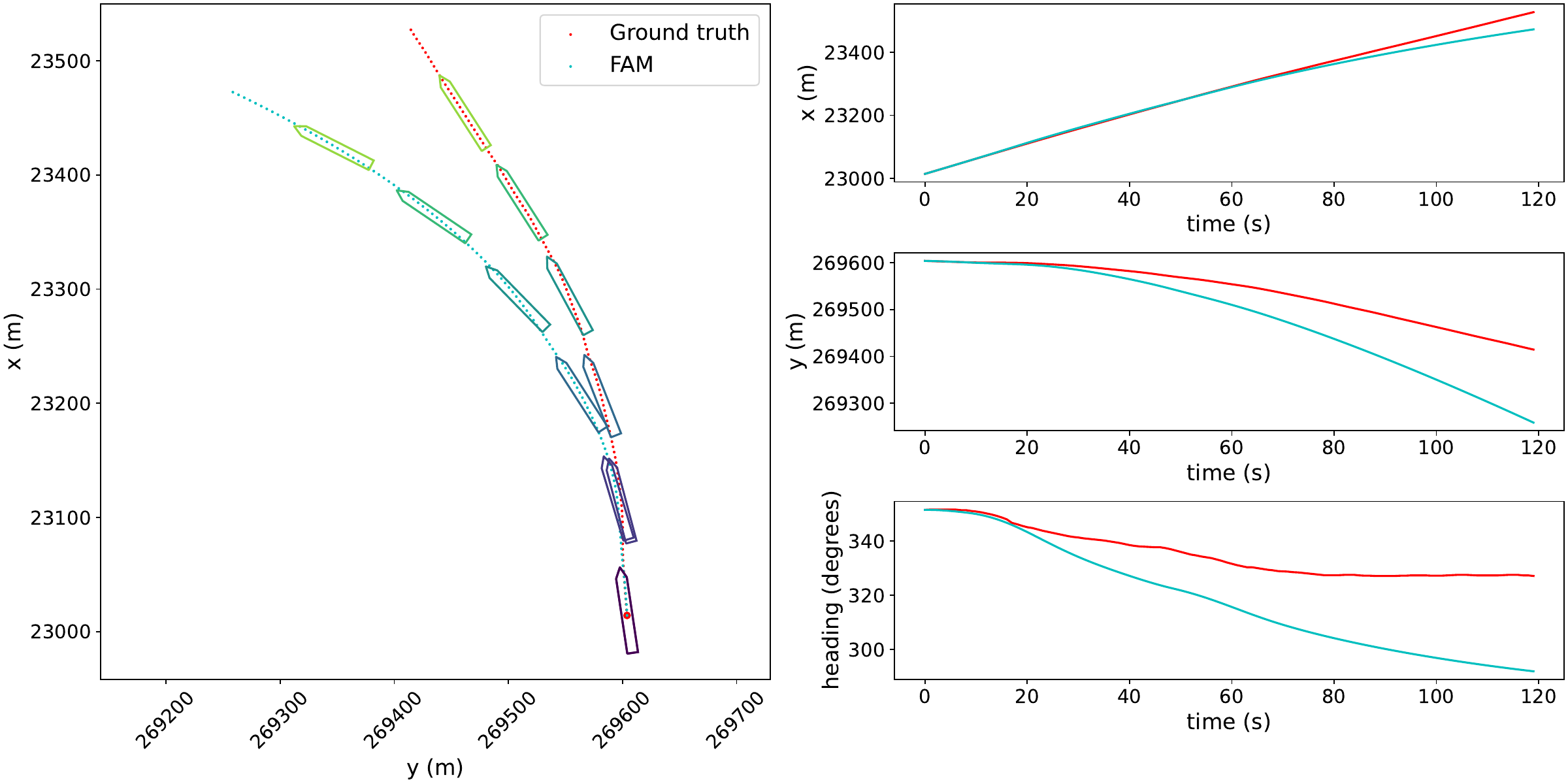} \vspace{-2mm}
    \caption{Comparison of FAM vs Real world data. Sub-optimal scenario 2. The turning of the vessel widens the difference of FAM vs actual vessel trajectory in almost all dimensions.}
    \label{fig:good6}
\end{figure}

For this evaluation phase, we utilized data from several dozen voyages of the SUZAKU vessel. We then breakdown each voyage to 2-minute trajectories. To illustrate the model's performance, we present two trajectory examples where FAM performs exceptionally well in~\cref{fig:good1,,fig:good2}, two examples with acceptable accuracy in~\cref{fig:good3,,fig:good4}, and two examples where the predictions of FAM are sub-optimal in~\cref{fig:good5,,fig:good6}. Further, the values of $cVDM$, which were used to categorize the trajectories are reported in~\Cref{table:errors_real_world}.

The examples with the good performance (optimal category) highlight the model's capability to accurately predict the ground truth trajectories despite the inherent challenges of real-world conditions. These scenarios include typical operational conditions and show that FAM achieves accurate trajectory predictions.

The acceptable results (satisfactory category) demonstrate that while the model may not be perfect, it still performs within a reasonable margin of error. These trajectories show some deviations but remain largely consistent with the ground truth trajectories. Please note that these deviations are mostly towards the end of the trajectories, hence they have a minimal role.

It should be noted that the vast majority of the real-world trajectories compared against the predictions of our model did actually fall into optimal or satisfactory category. This enhances our confidence that FAM can be further utilized within future control problems towards vessel autonomy.

The sub-optimal examples are particularly instructive as they reveal the limits of the model. These trajectories diverge significantly from the ground truth trajectories, suggesting areas where FAM could be improved. Analyzing such instances is critical for refining the model and addressing its weaknesses, ensuring continuous improvement. 

More specifically, it seems that FAM is more likely to diverge from the ground truth trajectories when the rudder angle values are approximately $10^\circ$. We have identified that, for rudder angles close to that range, the force of the rudder is not that large and, consequently, the coupling of the sway with the yaw rate of turn might dominate and dictate the predicted motion of the vessel.
It is evident that this has a more apparent effect on the predictions of FAM than on the actual vessel, hence this guides FAM to turn more (see~\Cref{fig:good6}).   

\noindent However, when we consider large rudder angles, \textit{e.g.} $35^\circ$ which produce large rudder forces, FAM is able to attain better predictions. This can be potentially explained as rudder forces can be more prominent than the sway - yaw rate of turn coupling. This phenomenon, which as already mentioned is very instructive, will be revisited in future work as it is an area of research on its own.

These results underscore the robustness of FAM in diverse real-world conditions. The consistency in the optimal performing examples shows that the model can handle typical navigational scenarios effectively, but struggles a bit more to achieve high accuracy in transient states, where the rudder angle is dynamically altered (see~\Cref{fig:good6}).

\vspace{-3mm}

\dottedsection{CONCLUSIONS}
\label{sec:conclusions}
\vspace{-2mm}
A physics-based motion model of a ship was presented, which accounts for forces influencing a vessel's motion. The model uses 3D dynamics for surge, sway, and yaw to predict position and velocity over time, incorporating the vessel's hydrodynamics and environmental conditions such as wind, waves, and currents.

Tailored for an 83-meter containership, the model was validated against state-of-the-art models and real-world data from the SUZAKU vessel. Multiple distance measures were used to compare the actual vessel's trajectory with the model's predictions, demonstrating promising results in terms of accuracy and reliability. 

Comparing FAM's predictions with MMG model turning tests yielded nearly identical trajectories, reinforcing the robustness and accuracy of our approach. The positive outcomes from real-world data evaluations further confirm the model's practical applicability in various maritime navigation scenarios.

FAM's accuracy in real world scenarios with very small rudder angle values is good, resulting in very close to reality predictions. However, FAM's accuracy may diminish in scenarios involving slight turns, which could happen due to unmodeled phenomena. Finally, we show that the custom Vessel Distance measure ($cVDM$) is an informative measure that can be very useful in real-world scenarios and could alleviate practitioners from the tedious visual inspection of trajectories.

Validating this physics-based model against real-world data marks a significant step in ship dynamics simulations, laying the groundwork for autonomous navigation, operational optimization, enhanced safety, and a more sustainable maritime future.

The ultimate goal is to develop a model that can adapt to real-world data, improving its generalization and capturing dynamic phenomena such as fouling, which affects a ship's maneuverability over time. As future work, we will consider implementing a data-driven analytical model to complement the current approach aiming to enhance the model's ability to address these challenges and improve its predictive accuracy. Also more work could be done in exploring quantitative, robust and scalable validation methods.
\vspace{-3mm}

\dottedsection{ACKNOWLEDGMENTS}
\vspace{-3mm}
The authors are thankful to Mr. Chatzigeorgiou Symeon, (DeepSea Technologies), as well as to all people and organizations involved, for the insightful and helpful discussions and the support for this work. 
\vspace{-4mm}

\printbibliography
\end{document}